# THE ROLE OF $C_2$ IN NANOCRYSTALLINE DIAMOND GROWTH

J.R. Rabeau[a*], Y. Fan[b], P. John[a], and J.I.B. Wilson[a]

a) School of Engineering and Physical Sciences, Heriot-Watt University, Edinburgh, Scotland, EH14 4AS

b) Department of Applied Physics and Electronic & Mechanical Engineering, University of Dundee, Scotland, DD1 4HN

*present address: Micro Analytical Research Centre, School of Physics, The University of Melbourne, Melbourne, Australia, 3010

E mail: jrabeau@physics.unimelb.edu.au

## ABSTRACT

This paper presents findings from a study of nanocrystalline diamond (NCD) growth in a microwave plasma chemical vapour deposition (CVD) reactor. NCD films were grown using $Ar/H_2/CH_4$ and $He/H_2/CH_4$ gas compositions. The resulting films were characterised using Raman spectroscopy, scanning electron microscopy and atomic force microscopy. Analysis revealed an estimated grain size of the order of 50 nm, growth rates in the range 0.01 to 0.3 μm/h and $sp^3$ and $sp^2$ bonded carbon content consistent with that expected for NCD.

The $C_2$ Swan band (d $^3\Pi_g$ ↔ a $^3\Pi_u$) was probed using cavity ring-down spectroscopy (CRDS) to measure the absolute $C_2$ (a) number density in the plasma during diamond film growth. The number density in the $Ar/H_2/CH_4$ plasmas was in the range 2 to 4 x $10^{12}$ $cm^{-3}$, but found to be present in quantities too low to measure in the $He/H_2/CH_4$ plasmas. Optical emission spectrometry (OES) was employed to determine the relative densities of the $C_2$ excited state (d) in the plasma.

The fact that similar NCD material was grown whether using Ar or He as the carrier gas suggests that $C_2$ does not play a major role in the growth of nanocrystalline diamond.

# 1. INTRODUCTION

In recent years, nanocrystalline diamond (NCD) has been identified as a potentially useful material for many applications such as wear-resistant coatings, MEMS and electron field emitters. NCD films are composed of diamond grains of the order of 50 nm, and under certain conditions display smooth morphology. In order to fully exploit NCD, a greater understanding of the chemical vapour deposition (CVD) growth process is essential to produce material suitable for specific applications. In particular, identification of the primary growth species would be useful for optimisation of the process parameters. Several research groups have employed spectroscopic techniques to measure the predominant species present during growth as a function of process parameters, especially under conditions of high argon concentration, typically argon/hydrogen/methane gas mixtures. Correlation between materials grown and species present in the growth environment is an important step in identifying the primary growth species.

Gruen and co-workers[1,2,3,4,5] have been major contributors to NCD research over the past 10 years. Preliminary work[2] demonstrated the growth of NCD using $C_{60}$ in an $Ar/H_2$ microwave discharge. Strong $C_2$ Swan band optical emission was observed using OES and from this it was proposed that $C_2$ may be the growth species for NCD. Ab-initio calculations supported the hypothesis that $C_2$ could insert directly into the C-H bonds which terminate the growing diamond surface. Similar results were obtained using $Ar/H_2/CH_4$ gas compositions in a microwave plasma using low $H_2$ (<10%) and high Ar (>90%) plus a constant $CH_4$ proportion (~1%). However *quantitative* data on the species present in the plasma is lacking together with characterisation of resulting diamond films. Correlation in this context may elucidate precisely what processes contribute to NCD film growth.

Several non-intrusive techniques are available for studying plasma chemistry. The simplest is OES which measures the spectral emission of the plasma and provides relative populations of excited state species. OES suffers from the inability to easily calibrate excited state emission intensities with absolute concentrations in the ground state. $Ar/H_2/CH_4$ plasmas have been shown to produce strong $C_2$ Swan band emissions ($C_2$ d-a) at around 516 nm, giving the plasma an emerald green colour. Laser induced fluorescence (LIF) is a non-invasive spectroscopy technique which can directly probe ground state levels and has been employed widely to study CVD chemistry. LIF has the advantage of high spatial resolution combined with negligible background effects. Careful calibration with a known quantity is required however to extract absolute concentrations. Absorption spectroscopy, where the change in intensity of light is measured after passing through a sample, is the simplest way to measure absolute concentrations (Beer-Lambert law). Because the technique may involve measuring small changes in intensity of the probe beam, it is limited by the inherent fluctuations of the light source. One way to improve on this limitation is to increase the path length of the beam through the sample by, for example, using a White-cell arrangement. Goyette *et al.*[6] used absorption spectroscopy in parallel with OES to show a correlation between the excited $C_2$ (d) state and the lower $C_2$ (a) state while changing parameters in a microwave plasma.

Cavity ring-down spectroscopy (CRDS) is a direct absorption technique which is an effective, non-invasive probe allowing high-sensitivity measurement of the absolute concentrations of species in hot filament, flame and plasma environments. Early work was done by Wahl *et al.*[7] who used CRDS to study the methyl radical ($CH_3$) in a resistively heated tungsten hot-filament reactor for diamond deposition. The methyl radical may be an important precursor to diamond formation and is often considered to be the growth species in $CH_4/H_2$ plasmas.[8,9] Quantitative number density spatial profiles for $CH_3$ were measured at different distances from the hot-filament and at different temperatures. Further studies[10] showed good agreement between experimental data and two-dimensional models of the system. Lommatzsch et al.[11] used CRDS to study the $A^2\Delta$ - $X^2\Pi$ transition of CH at 431 nm in a diamond forming hot-filament reactor. Rotational temperature and concentration profiles were measured with respect to distance from the filament. Scherer et al.[12] used infrared (3200 cm$^{-1}$ ~ 3.1 μm) CRDS in conjunction with coherent anti-Stokes Raman spectroscopy (CARS) to determine absolute methyl radical concentrations in low pressure methane-air flames. They achieved a fractional absorption sensitivity of 3 – 10 ppm/pass. Absolute concentrations of CN, CH and $CH_2O$ were measured using CRDS and LIF in low pressure pre-mixed flames by Luque et al.[13,14]. The combination of spatially resolved LIF measurements and highly sensitive absolute number densities from CRDS produced quantitative two-dimensional images of the flame. The absolute ground state population of the methylidine radical (CH) was measured by Engeln et al.[15] at ~430 nm in an Ar/$C_2H_2$ expanding arc plasma. The density was measured as a function of arc current and acetylene flow rate. Methylidine trends showed that it likely plays a minor role in the deposition of hydrogenated amorphous carbon.

CRDS has been used for the detection of $C_2$ radicals in the $A^1\Pi_u$ state and the $a^3\Pi_u$ state and most recently $X\ ^1\Sigma_g^+$. Staicu et al.[16] measured the absolute concentration of the $A^1\Pi_u$ state of $C_2$ in an atmospheric oxy-acetylene flame. The authors estimated the temperature distribution and applied the Abel inversion technique to convert the CRDS measured column densities to absolute number density profiles. Peak concentrations ranged from $8 \times 10^{14}$ to $2.5 \times 10^{15}$ m$^{-3}$. Absolute CN radical concentrations were measured in the same oxyacetylene flame during diamond deposition. Column density measurements showed that nitrogen addition influenced mainly the central region of the flame (within a radius of ~2.7 mm). CH was also measured[17] and a correlation between local growth rate and CH concentration profiles were established. CH was concluded to play an important role in the formation of an annulus on the film, which displayed enhanced diamond growth.

Benedikt and co-workers[18] studied the species C, $C_2$, CH, and $C_2H$ in a remote Ar/$C_2H_2$ expanding thermal plasma. The $C_2$ ($a^3\Pi_u$) radical behaviour was measured as a function of $C_2H_2$ flow and the a-C:H films were characterised. The plasma chemistry under these conditions was dominated by charge exchange (Ar and $C_2H_2$) and dissociative ion-electron recombination. The first demonstration of CRDS applied to microwave plasmas was published by John *et al.*[19] who measured $C_2$ as a function of methane content and correlated the $a^3\Pi_u$ state in absorption with the $d^3\Pi_g$ state in emission (using OES). A linear relationship between the CRDS and OES signals was demonstrated. Wills *et al.*[20] measured $C_2$ and CH in a dc arc jet reactor. They used the spectral data in conjunction with spectral simulations to determine the rotational temperature (3300 K) and absolute number densities ($6 \times 10^{12}$ to $1.5 \times 10^{13}$ cm$^{-3}$) of $C_2$ ($a^3\Pi_u$). These experiments were done under conditions favourable for diamond growth although no information was provided on the nature of the films.

The low lying $C_2$ ($a^3\Pi_u$) electronic state is only 716 cm$^{-1}$ above the absolute ground state (X $^1\Sigma_g^+$). At the temperatures expected in the plasma CVD process environment, the metastable a-state should therefore be highly populated. $C_2$(X) has much faster reaction rates with molecular hydrogen and hydrocarbons than $C_2$(a),[21] and it was therefore thought that the $C_2$ (a) state would be more densely populated than $C_2$(X). However, it was recently shown using CRDS[22] in an arc jet reactor that the $C_2$(X) relative to $C_2$(a) population is very close to that predicted by a Boltzmann distribution indicating thermal equilibrium. This result suggested a negligible depletion of $C_2$(X) with respect to $C_2$(a), or intersystem crossing between the two states which are rapid enough to maintain thermal equilibrium and compensate for the enhanced reactivity of $C_2(X)$.

To date most NCD work has concentrated on the effects of progressively adding Ar to the growth process (whilst reducing the $H_2$ content). We present a study which compares separately the addition of argon and helium to the growth process, and CRDS measurements examine the absolute number density of $C_2$ within the plasma during NCD growth. With the addition of Ar, there was an abundance of $C_2$ ($\sim 10^{12}$ cm$^{-3}$) and with the addition of He, the quantity of $C_2$ was below the detection threshold of our spectrometer ($< 10^9$ cm$^{-3}$).

The work presented here raises questions regarding the proposition that $C_2$ is the key growth species for nanocrystalline diamond. Based on the evidence obtained from experiments, with Ar and He as diluents, it appears that NCD films can be grown under conditions where the $C_2$ concentration is extremely small.

## 2. EXPERIMENTAL

### A. CVD reactor

The stainless steel reactor chamber had four CF 250 ports on each side, one connected via a gate valve to a load lock. A 19 cm long and 19 cm diameter quartz cylinder with microwave shielding mesh was mounted on top of the main chamber and provided the main reaction volume. Six symmetrical CF70 ports were added via quartz extension tubes fused to the quartz cylinder. These ports were arranged axially, providing the capability for cavity ring-down spectroscopy measurements of up to three species simultaneously. The roughing and backing pump was a two stage E2M40 HV rotary pump (Edwards) and the turbomolecular pump was an EXC 300 controlled EXT 501/160 CF (Edwards). Pressure was measured using an active Pirani gauge APG-M-NW16 AL (Edwards) and a Barocel 600 capacitance manometer (Edwards). A throttle valve model MDV-015S06 (Tylan General) and Model 80-2 controller (Vacuum General) was used to control the chamber pressure.

A coaxial blade reactor (CBR) which employed a novel microwave power applicator (DILAB Ltd) was used to grow the films discussed in this paper. The power applicator had 12 blades distributed evenly around a 19 cm diameter aluminum plate. The blades were 2.5 cm wide and extended 6 cm downwards from the bottom edge of the plate (see FIG. 1).

The microwave supply and magnetron was a Muegge model MW-GPERE 3327-5K-02, with a frequency of 2.46 GHz, 6 kW max. The microwaves were guided through a rectangular, fan cooled waveguide to the microwave applicator. The waveguide was equipped with three tuning stubs and a tuning plunger for optimal mode matching into the microwave cavity.

The 'low flow' gases ($H_2$ and $CH_4$) were controlled with Ultraflo Massflow controllers (Vacuum General) capable of 0-100 standard cubic centimetres per minute (sccm) flow rates. The 'high flow' rate gases (Ar and He) were controlled using Tylan FC2901 mass flow controllers (Millipore) capable of 0-1 standard cubic litres per minute (sclm) flow rates.

$H_2$ and $CH_4$ were premixed after the mass flow controllers (MFC) and flowed into the chamber through ¼ inch stainless steel piping and distributed around the inside perimeter chamber through a specially machined stainless steel ring with small axially spaced holes. Ar and He were fed through ¼ inch ports added to the quartz extension tubes. The high inert gas flow kept the volume within these tubes free of particulates, which drastically affected the cavity ring-down spectroscopy measurements when present.

During growth, the substrate temperature was measured in the reactor using a one-colour optical pyrometer mounted to the top of the reactor with the alignment optics focused on the silicon substrate. The emissivity of the silicon substrate was taken to be 0.62 under the estimated temperature conditions (~ 500 °C).

Optical emission spectra (OES) were collected using a PC-controlled, grating based, optical spectrum analyser (Monolite 6800). A fibre optic cable attached to a small view port on the plasma reactor guided the light to the input slit of the spectrometer.

**B.     Growth and Analysis**

Substrates were nucleated by manually polishing an entire 4 inch wafer with 0.5 to 0.75 μm diamond powder (DeBeers Industrial Diamond). The wafer was then cut into 2 x 2 cm squares using a diamond-scriber and cleaned with acetone, methanol, Decon solution and deionised water.

Growth experiments in the CBR focused on the change of gas composition, specifically the increase of inert gas proportion (Ar or He) added to $CH_4$ and $H_2$. Diamond films were analysed for trends in surface morphology, growth rate, diamond/graphite content and crystallite size. Raman spectroscopy, atomic force microscopy (AFM) and scanning electron microscopy (SEM) were used.

The following parameters were maintained constant for all experiments unless stated otherwise: 2.1 kW microwave power, 90 Torr (12 kPa) total pressure, 7.5 hr growth time, 480 sccm total flow. Table I summarises the substrate temperature and gas flow rates for each sample run.

Raman analysis was performed at Leeds University, UK, with a Renishaw RM1000 spectrometer with an intracavity frequency-doubled 244 nm continuous wave argon ion laser (Innova 300 FreD, Coherent Inc., Santa Clara, CA). All spectra were obtained with 1 mW laser power through a 40x objective lens.

The scanning electron microscope (SEM) was an Hitachi 2700 SEM, which detects secondary electron emission from the sample surface. The system was operated with 10 keV electron energy. Atomic force microscope (AFM) images were collected using a Dimension 3000 Atomic Microscope (Digital Instruments, USA).

## C. Cavity ring-down spectroscopy

### 1. CRDS theory

CRDS measures the rate of decay of a laser pulse passing back and forth through a sample contained within an optical cavity. A pulse of light is injected into a high-finesse optical cavity consisting of two axially aligned high reflectivity mirrors M1 and M2. In this instance, most of the light will be back-reflected, but a small fraction will leak into the cavity and proceed to M2. When this 'attenuated' pulse reaches mirror M2 within the cavity, again, most of the light will be back-reflected toward M1 and a small amount will leak out of the cavity. A detector placed on the other side of M2 to detect the fraction of light leaking out after each round-trip of the pulse within the cavity will show that the pulse intensity decays according to the Beer-Lambert law:

$$I = I_0 \exp(-t/\tau) = I_0 \exp(-kt)$$  **Equation (1)**

Where I is the initial intensity, $I_0$ the instantaneous intensity, $k$ is the decay rate and $\tau$ is the time for $I = I_0/e$. The resulting decay-curve is known as a "ring-down" curve. The pulse will ring-down faster if there is an absorber in the cavity, and the spectrum of the absorber can be defined using a tuneable laser to scan the wavelength over a spectral region of interest and measuring the decay rate at different each wavelength point.

The change in the ring-down rate coefficient, $\Delta k$ (off and on-resonance with a spectral feature in the cavity – in this case $C_2$ (a-d), is related to the absorbance, $\alpha$, by the following:

$$\alpha = \frac{L\Delta k}{lc}$$  **Equation (2)**

Where *L* is the RD cavity mirror spacing, *l* is the estimated path length through the plasma and c is the speed of light. *L/l* accounts for the fact that the species is not homogeneous over the entire cavity length, but confined to the plasma volume. In these experiments the plasma volume was defined as the volume where emission was visible by eye and *l* was measured by comparing the plasma diameter with the known diameter of the substrate holder. The integrated absorption coefficient is related to the $C_2$ number density by the following equation:[20]

$$\int_{line} \alpha_v dv = \frac{\lambda^2}{8\pi c} \frac{g_d}{g_a} [C_2(a)] A_{00} p \frac{1}{Q_v} \qquad \textbf{Equation (3)}$$

Where, $\lambda$ is the transition wavelength, $g_d = g_a = 3$ are the electronic degeneracies,[23] $[C_2(a)]$ is the number density of $C_2$ in the a-state over all vibrational levels, and $A_{00}$ (7.21 ± 0.30 × $10^6$ $s^{-1}$) is the Einstein A coefficient for the *(0,0)* band taken from Wills *et al.*[20] which includes the fluorescence lifetime ($\tau$ = 101.8 ± 4.2 ns, d $^3\pi_g$, $v^1$ = 0) and the Franck-Condon factor ($Q_{00}$ = 0.7335). The factor *p* is the fractional contribution of the integrated line to the total *(0,0)* band oscillator strength at a given rotational temperature ($T_{rot}$). Spectral simulations and fitting were performed using PGopher[24,25] to estimate $T_{rot}$ and calculate the *p* factor. PGopher simulations include Boltzmann factors and line strengths and are thus useful for simple computation of the temperature dependent *p* factor. The final factor, $Q_v$, is the vibrational partition function, which accounts for the Boltzmann distribution of populations in higher vibrational levels. The integrated absorption coefficient, $\alpha_v dv$, was measured by fitting a Gaussian curve to the spectral feature and determining its area.

2.   *CRDS experimental*

The cavity ring-down spectrometer consisted of a wavelength tuneable probe laser (Continuum Surelite Nd:YAG SLI-10 pumped dye laser Sirah Laser-und Plasmatechnik

GmbH, 2400 lines/mm grating) operated with coumarin 307 radiant dye for the measurement of the $C_2$ Swan band (band head at 516.5 nm).

The pump laser typically ran at 10 Hz giving a frequency-tripled 355 nm output and 7 – 10 ns pulse duration. The attenuated probe beam (dye laser, 1-2 mJ after neutral density filtering) was passed through a pair of telescoping lenses, which provided a beam diameter of 2mm.

The probe laser pulses were injected into a 70 cm high-finesse optical cavity, the axis of which crossed the centre of the plasma. The mirrors were 6m radius of curvature and 2 cm in diameter with a factory reflectivity of >99.997% centred at 520 nm (Los Gatos Research). The mirrors were housed in 2.75 inch vacuum flange mounts with three external screw actuators and mounted with Viton gaskets to the vacuum chamber and protected from possible contamination by a continuous flow of argon or helium.

The beam was steered into the ring-down cavity with several silver coated mirrors (Edmund Optics, R > 99%, 450 – 1200 nm).

Light was detected at the output of the end mirror with a photomultiplier tube (Hamamatsu R955 side-on PMT, 160 – 900 nm responsivity) biased between 500 and 600 V (EG&G Ortec 456 HV power supply). The signal was sent to a LeCroy 9361 digital oscilloscope (300MHz). Background plasma emissions were blocked with a pair of long and short pass optical filters (to allow ~450 to 550 nm transmission). The ring-down curve was captured and sent via GPIB to a PC and processed with LabVIEW software (National Instruments).

## 3. RESULTS AND DISCUSSION

### A. Plasma diagnostics

OES spectra were measured for the experimental conditions listed in Table 1. FIG. 2 shows typical OES spectra for Ar and He containing plasmas. $C_2$ (d) was observed in abundance with Ar addition, but in very low levels with He addition. Conversely, hydrogen Balmer emission lines were very intense in the He plasmas, but very weak in the Ar plasmas.

CRDS was employed to measure the rotational temperature ($T_{rot}$) and absolute number density of $C_2$ by relating the ring-down coefficients for different spectral lines in the $C_2$ Swan band to the absolute number density according to Equation (3). The average $T_{rot}$ for all conditions employing $Ar/H_2/CH_4$ gas mixtures was 3010 K ± 340 K (± 11%). This temperature was used for the calculation of the absolute number density of $C_2$ in the plasma. The number density was measured between 85 and 99 % Ar addition. FIG. 3 shows the change in $C_2$ concentration with respect to the Ar concentration. With the addition of He to the gas mixture, the $C_2$ concentration was unmeasurable, indicating levels below the detection threshold of the spectrometer.

B.   **Diamond film analysis**

1.   *Diamond and non-diamond carbon*

Raman spectra from films grown with either Ar or He dilution are shown in FIG. 4. UV Raman spectroscopy assessed the *sp³* and *sp²* bonded carbon content, which is an indication of, respectively, diamond and graphite in the material. Deconvolution of the spectra using Lorentzian fits revealed three significant peaks. One at ~1333 cm$^{-1}$ was assigned to diamond[26] (*sp³* bonded carbon) and one slightly broader peak centred at ~1595 cm$^{-1}$ was assigned to graphite (*sp²* bonded carbon, G-band).[27,28,29] The third peak appeared between 1300 and 1350 cm$^{-1}$ and was assigned to *sp²* bonded carbon (D-band).[27,30] A fourth, broad curve was also included in the fit which accounted for the broad background luminescence.

A small, broad peak is apparent in some of the spectra, centred at ~ 1440 cm$^{-1}$. This peak has been associated with nanocrystalline diamond and is thought to be due to the presence of transpoly acetylene.[28] Another small peak is visible in some of the spectra centred at ~1555

cm$^{-1}$. This has been assigned to $sp^2$ carbon.[27,28,29,31] A typical deconvolution of the 244 nm Raman spectra is shown in FIG. 5.

FIG. 6 shows the integrated areas of the diamond and 'graphite' related peaks ($sp^3$ and $sp^2$ respectively) for both Ar and He dilution in the microwave plasma. The plot reveals that the $sp^3 / sp^2$ carbon proportion is constant up to 97% inert gas, where it increases in the Ar case and decreases in the He case. This may be attributed to the non-continuous films under these conditions, with the laser tending to sample local fluctuations in the material, not necessarily representative of the bulk properties.

Raman analysis of films grown with Ar and He dilution revealed the presence of peaks and relative intensities consistent with that expected for nanocrystalline diamond.[28,30,31] Broadened and shifted diamond lines (1333 cm$^{-1}$) in all the films suggest a reduction in the grain size and high compressive stress[32] from the phonon confinement model.[26,33] This is expected to occur in nanocrystalline diamond and has been observed previously in nanocrystalline diamond films.[26,33]

The development of a shoulder on some of the diamond lines in the 244 nm Raman spectra may be due to the D-band, which is attributed to $sp^2$ bonded carbon. The D-band was clearly visible after deconvolution of the UV Raman spectra. This feature was also observed in nanocrystalline films analysed by Ferrari *et al.*[28]

*2.     Growth rate*

Cross sectional SEM micrographs were taken of the deposits and the average film thickness measured. Two sample micrographs of films grown in Ar and He are shown in FIG. 7. Some films were not continuous, and in this case the thickness was determined based on the average height of the isolated deposits. The Ar grown films were uniform and continuous across the length of the substrate under all conditions except for 99% Ar. The He grown films were continuous but rough in comparison to the Ar films and consisted mainly of large "ballas" structures.

The growth rates for these films are plotted as a function of inert gas addition in FIG. 8. The growth rate was shown to remain relatively constant over time from measurements made on films grown for 1, 3, 5 and 7.5 hours. The error in measuring the growth rates was estimated to be ± 10 %. Growth rates between Ar and He films were very similar, suggesting the growth mechanism is the same with Ar and He dilution.

### 3.   Crystallite size (AFM)

The AFM analysis yielded high resolution images and immediately revealed that the deposits were composed of clusters of tiny grains. Nanocrystalline diamond has been defined as being polycrystalline in nature consisting of diamond grains in the range 50 to 100 nm.[34] With respect to grain size criterion, films grown with 85 to 99% Ar and He addition are consistent with the grain size expected for NCD. AFM images are shown in FIG. 9 for Ar and He diluted films (90% only). Over the range 85 to 99% Ar and He, the grain size appeared to be uniform. The mean external grain size was estimated from the AFM images to be ~50 nm in the Ar films and ~53 nm for the He films.

### 4.   Surface roughness

The surface roughness of the films, defined as the RMS of the distribution of "heights" for each point on the scan, was calculated from the AFM data over 10 μm areas and is plotted in FIG. 10. There is clearly a morphological difference between He and Ar grown films, a feature which is also evident in the SEM images. The possible reasons for this are discussed later.

5.  *Summary of Results*

In summary, NCD films which were very similar in $sp^3$ and $sp^2$ carbon content (assessed using Raman spectroscopy), grain size (assessed using AFM) and growth rate (assessed using SEM) were grown using two different gas mixtures: Ar or He diluted $H_2/CH_4$. In the case of Ar dilution, the $C_2$ radical density was measured to be ~ $10^{12}$ cm$^{-3}$ using CRDS and showed intense Swan band emissions in the OES spectra, whereas in the case of He dilution, the $C_2$ radical density was too low to measure using CRDS and showed only a small emission peak in the OES spectra.

Using He as a diluent, we expected to find similar plasma characteristics to Ar containing, particularly with respect to $C_2$ concentration. However, $C_2$ was present in quantities below the sensitivity of our cavity ring-down spectrometer (estimated to be ~$10^9$ cm$^{-3}$) and at least a factor of 1000 less than that present in the Ar containing plasmas. This finding was confirmed by OES measurements which showed very weak emission from the $C_2$ (d) state. It has previously been shown using CRDS and OES that the $C_2$ (d) and $C_2$ (a) states correlate linearly as a function of plasma condition in an $Ar/H_2/CH_4$ plasma.[19]

The fact that similar NCD films were grown both in the presence and in the absence of $C_2$ brings into question whether $C_2$ contributes significantly to the growth process.

C.  **Does $C_2$ play a role?**

The abundance of $C_2$ in Ar containing plasmas can be attributed to the enhanced pathways available for $C_2$ production, arising from the presence of Ar* in the plasma where energy transfer with $C_2H_2$ and/or $C_2H$ forms $C_2$.[35] The first excited state of He (He*) has a higher energy (19.8 eV compared to 11.55 eV for Ar*) which arguably implies that it will be less populated that Ar*. Dissociation reactions with $C_2H_2$ and $C_2H$ (ex. Ar* + $C_2H_2$ → Ar + $C_2$ + $H_2$) will therefore be minimal

The growth rates measured for films grown in this study are in general agreement with previously published results, which typically indicate a decrease in growth rate with increased addition of inert gas (> 60%).[5,36,37,38,39] FIG. 11 presents a comparison of the findings of several studies by different research groups including our own. Zhou et al.[5] analysed the growth rate as a function of Ar addition with constant 1% $CH_4$, a balance of $H_2$ and a constant 800°C substrate temperature. Yang et al.[36] carried out a similar $Ar/H_2/CH_4$ study with a constant temperature of 900°C. Baranauskas et al.[37,38] compared growth rates with the addition of He and Ar with a constant 0.5% $C_2H_5OH$ and a balance of $H_2$. They maintained a constant substrate temperature of 855 °C by increasing the filament current in their HF reactor, which may have influenced the growth rate. Han et al.[39] increased the Ar content in a 2% $CH_4/H_2$, 1 kW microwave plasma and grew films at 800 °C to 900 °C substrate temperature. The results from this work are also included in FIG. 11 for comparison. The plotted growth rate magnitudes shown in FIG. 11 were estimated from the published plots of growth rate versus vol% inert gas and do not necessarily indicate the precise magnitudes. With respect to $C_2$ content, Han et al.[39] and Zhou et al.[5] reported an increase in $C_2$ content in the gas-phase as measured by OES.

FIG. 11 shows that the growth rate is influenced by inert gas addition and a reduction in H/C, and can often cause an increase in growth rate with ~ 0 to 50% dilution, and a decrease with 50 to 99% dilution.

Despite the general similarities in conditions and results, there is disagreement in the proposed mechanisms responsible for the trends in the growth rate:

Yang and co-workers[36] suggested that the initial drop in growth rate during their experiments (0 – 10% Ar) was due to the decrease in H required for the $CH_3$ growth mechanism.[8] The subsequent increase in growth rate at 10% Ar was attributed to the $C_2$ mechanism[40] becoming dominant. The growth rate was said to reach a maximum where the $C_2$ density reached a 'steady state' and then decreased because of the continued dilution of active species by Ar.

Zhou *et al.*[5] maintained that, although $C_2$ continued to increase while the growth rate decreased (at >60 % Ar addition), $C_2$ was still the growth species. This was explained by calculations[41] that show there is a pathway for diamond production via $C_2$ insertion that includes atomic H addition to the double bond of one ethylene-like group, and has a lower activation energy than the pathway that does not include H. The direct $C_2$ insertion mechanism has an activation barrier of *ca.* 2 kcal/mol.

Baranauskas *et al.*[38] proposed a slightly different argument, suggesting that Ar effectively diluted the radicals at the filament surface and thus increased their lifetime. The increased lifetime allowed them to travel further from the filament and may have ultimately increased their concentration at the growing substrate surface, thus increasing the growth rate.

There are obvious difficulties in establishing a consistent model which accounts for the change in growth rate as a function of conditions. Attempts have been made to establish a balance between the two most likely growth species, $CH_3$ and $C_2$, over a range of conditions, however the arguments discussed above are complex.

FIG. 11b shows growth rates in the region 80 to 99% inert gas with the corresponding measured $C_2$ number density from the Ar containing plasmas superimposed. Aside from the trend of $C_2$ number density, which does not compliment the measured diamond growth rate, there is an obvious disparity with the He results in that the $C_2$ number density was too low for detection (at least 3 orders of magnitude less than in the Ar plasma).

We propose that $C_2$ cannot be the growth species for the formation of NCD. Firstly, because it simply is not present in quantities large enough[42] to contribute in the case of He dilution, and secondly, because when it is present (in Ar dilution) it does not correlate with the measured change in growth rate. The fact that the growth rates measured under both conditions are nearly identical suggests the same mechanism is at work.

The widely accepted growth species for diamond in $H_2/CH_4$ gas mixtures is the methyl radical, $CH_3$.[8,9] This model involves surface H-abstraction reactions, insertion of $CH_3$ and further H-abstractions to form new diamond layers. Nanocrystalline diamond can be grown under conditions of high $CH_4$ containing (> 10 %) $H_2/CH_4$ gas mixtures.[43,44,45,46] Hiramatsu *et al.*[43] reported a 50 nm grain size from diamond films grown in a microwave plasma at 55 Torr with 10% $CH_4$ in $H_2$. Incidentally, they measured a $C_2$ number density of ~ 2 × $10^{11}$ cm$^{-3}$ in the plasma using absorption spectroscopy. Haubner and Lux[46] refer to 'coarse ballas' diamond (nanocrystalline diamond clusters) grown in a hot filament reactor with 10% $CH_4$ in $H_2$. Chen and co-workers[44] found grain sizes of less than 100 nm on diamond films grown with 4 to 41% $CH_4/H_2$ microwave plasmas.

Whether using an inert gas diluent or pure $CH_4/H_2$ gas compositions, NCD seems to form in an H-depleted environment (reduction in H/C), perhaps where there is an enhanced mechanism for secondary nucleation. Evidence the fact that "ballas" or "cauliflower" diamond is often observed at the H-depleted perimeter of films grown in 1% $CH_4/H_2$ gas mixtures. If the $CH_3$ mechanism is indeed at play in our experiments, at some point the reduction in H/C would hinder the process and cause a decrease in the growth rate (from FIG. 11, this point appears to be at around 60% dilution). It is possible that the initial rise in growth rate is due to an enhancement in the mean free path of reactive species within the reactor. At ~60% dilution with inert gas (39% $H_2$ and 1% $CH_4$) the reduction in H is at such a level that H-abstracted surface sites become limited and the insertion of $CH_3$ radicals is inhibited. For comparison with gas mixtures used in this work, a 10% $CH_4/H_2$ gas mixture (such as that used by Hiramatsu *et al.*[43]) is equivalent to 90% Ar or He, 9% $H_2$, 1% $CH_4$.

A significant feature of the films is the difference in surface morphology observed in the SEM micrographs. The Ar grown films took on a smooth morphology, while the He grown films were rough. This may be attributed to enhanced re-nucleation of the growing film in the presence of $C_2$, which effectively increases the uniformity over the entire surface. In the absence of enhanced re-nucleation, the He films tended to grow in large clusters, perhaps around the original Si-surface nucleation sites. The difference in surface morphology may also be attributed to the difference in mass between Ar and He. It is plausible that heavier neutral and ionic Ar continually bombards the growing surface and maintains uniform smoothness. Ar may be more effective at surface smoothing that He because of its higher mass, and may therefore be essential for the growth of smooth nanocrystalline diamond films.

## 4. CONCLUSIONS

Films were grown in the microwave reactor with Ar and He dilution from 85 to 99% (in a balance of $H_2$ and 1% $CH_4$). Films grown with Ar dilution revealed a uniform and smooth morphology (21 to 43 nm RMS roughness) in contrast to films grown with He dilution which revealed rough, ballas-like morphology (600 to 61 nm RMS roughness). AFM analysis also revealed uniform ~ 50 nm grains in both Ar and He films.

The Ar and He grown films were very similar in carbon composition as shown by Raman spectroscopy. Raman spectra indicated the presence of NCD according to relative *$sp^3$* and *$sp^2$* carbon content and spectral features associated with NCD.

The growth rates were measured over this range using cross-sectional SEM and found to be similar in magnitude (0.3 to 0.01 μm/h). The growth rate monotonically declined with increased inert gas content. The corresponding $C_2$ number density measured during Ar dilution reached a maximum at 95% Ar and declined with further Ar dilution.

The species, $C_2$, does not appear to contribute as the growth species in the formation of nanocrystalline diamond in $Ar/H_2/CH_4$ and $He/H_2/CH_4$ containing microwave plasmas. Adding inert gas and reducing $H_2$ in $H_2/CH_4$ gas mixtures causes a decrease the atomic H/C ratio. Similar growth mixtures can be achieved with a high $CH_4$ content in $H_2/CH_4$ gas mixtures (>10%). Evidence for this is that similar 'ballas' or nanocrystalline diamond has been observed using pure $H_2/CH_4$ gas mixtures.[43,44,45,46] NCD has been shown to form independent of the presence of $C_2$.

## 5. ACKNOWLEDGEMENTS

The authors wish to acknowledge financial support from the Engineering and Physical Sciences Research Council (Technological Plasmas Initiative) UK. The authors would like to thank Dr. Alastair Smith and Inigo Mendieta at the Unversity of Leeds for the Raman analysis. J.R.R. wishes to thank Dr. Andrew Orr-Ewing and Dr. Jonathon Wills of the University of Bristol for their introductory guidance with practical aspects of CRDS and supplying a copy of PGOPHER for $C_2$ spectral simulations

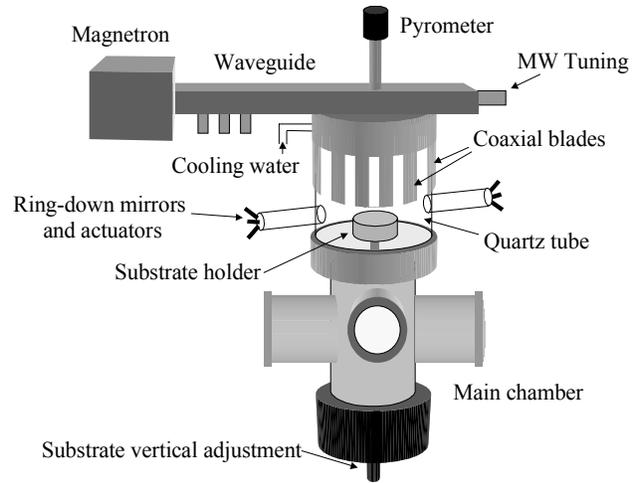

FIG. 1. J.R. Rabeau

|    | Sample | Ar/He (%) | $H_2$ (%) | $CH_4$ (%) | $T_{substrate}$ (°C) |
|----|--------|-----------|-----------|------------|----------------------|
| Ar | 6-016  | 85        | 14        | 1          | 515                  |
|    | 6-015  | 90        | 9         | 1          | 450                  |
|    | 6-010  | 95        | 4         | 1          | 450                  |
|    | 6-019  | 97        | 2         | 1          | 420                  |
|    | 6-020  | 99        | 0         | 1          | 400                  |
| He | 6-027  | 85        | 14        | 1          | 495                  |
|    | 6-029  | 90        | 9         | 1          | ~ 400 - 500          |
|    | 6-028  | 95        | 4         | 1          | ~ 400 - 500          |
|    | 6-030  | 97        | 2         | 1          | ~ 400 - 500          |
|    | 6-031  | 99        | 0         | 1          | ~ 400 - 500          |

TABLE I.  J.R. Rabeau

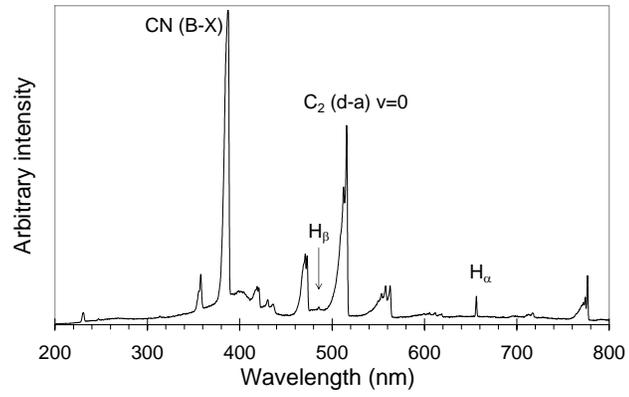

(a)

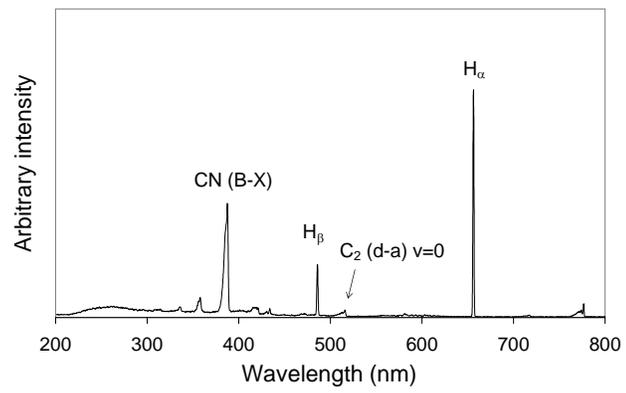

(b)

FIG. 2. J.R. Rabeau

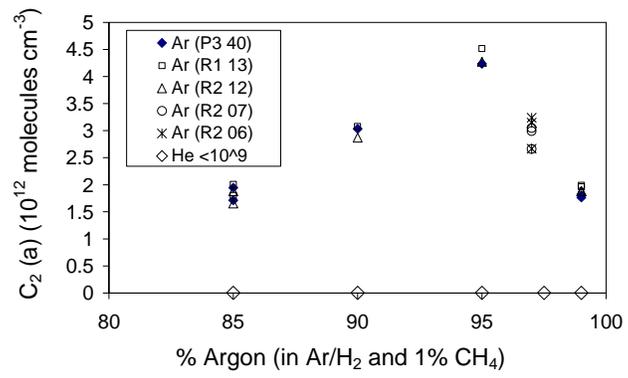

FIG. 3. J.R. Rabeau

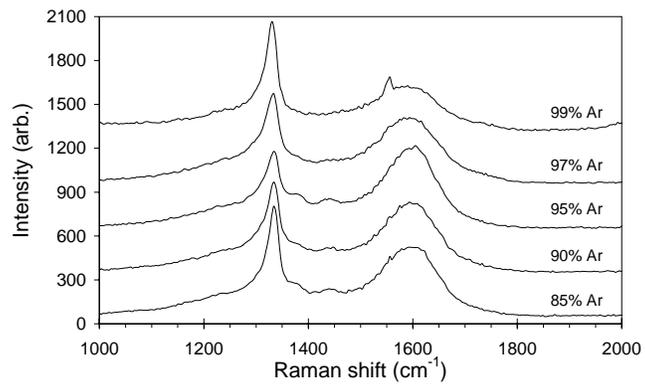
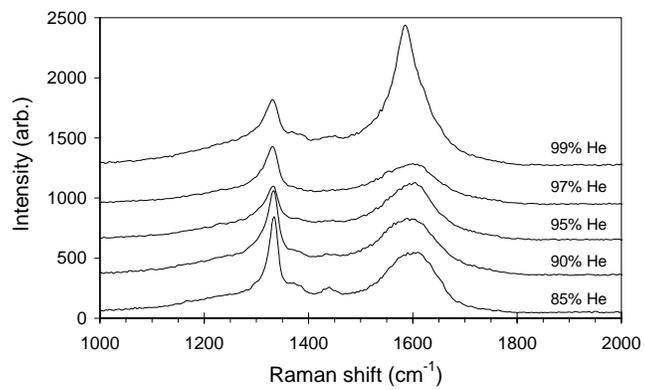

FIG. 4. J.R. Rabeau

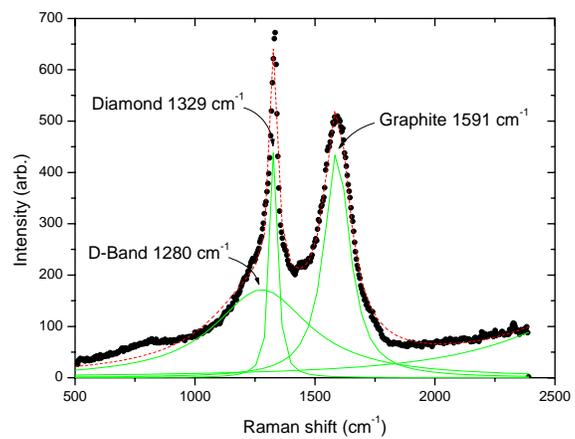

FIG. 5. J.R. Rabeau

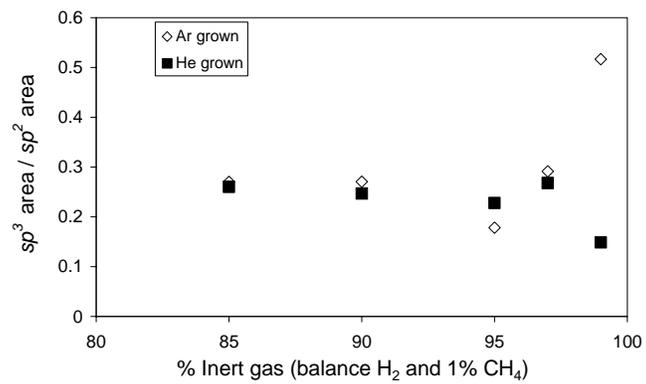

FIG. 6. J.R. Rabeau

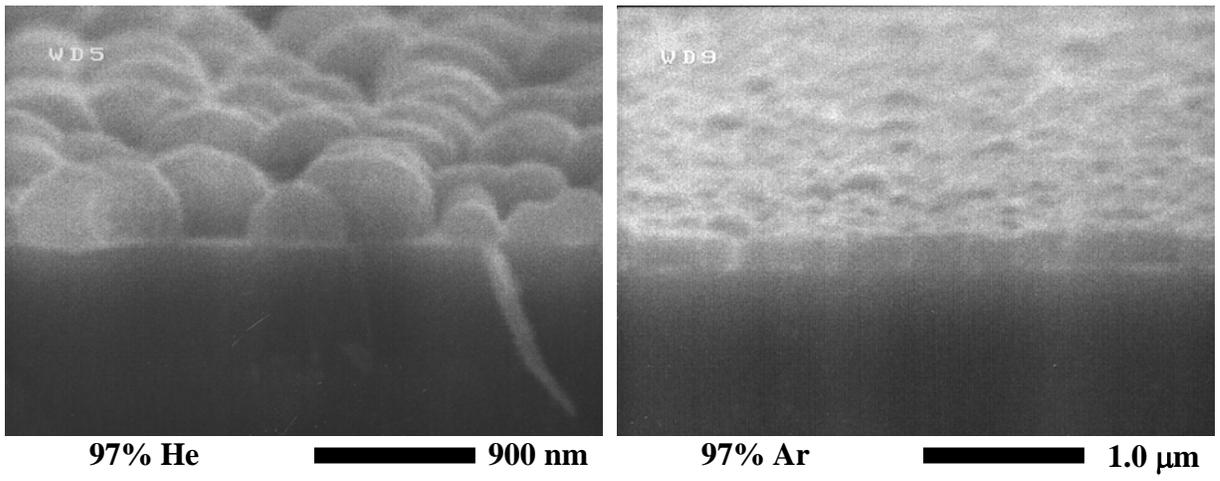

97% He    900 nm     97% Ar    1.0 μm

FIG. 7. J.R. Rabeau

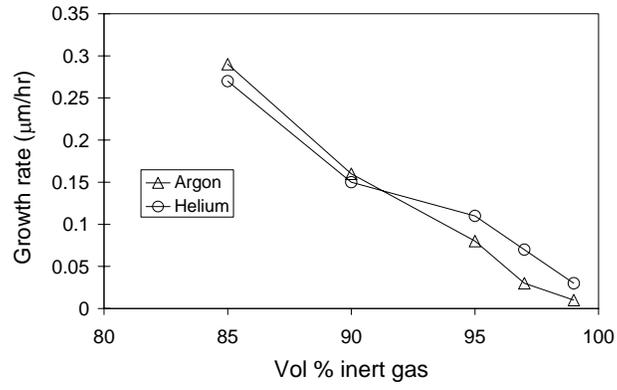

FIG. 8. J.R. Rabeau

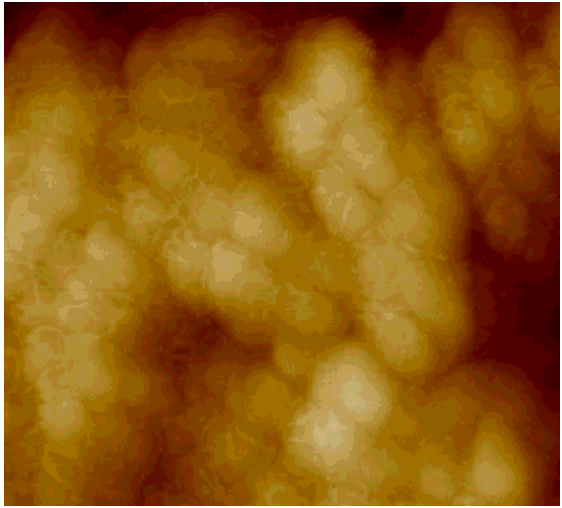 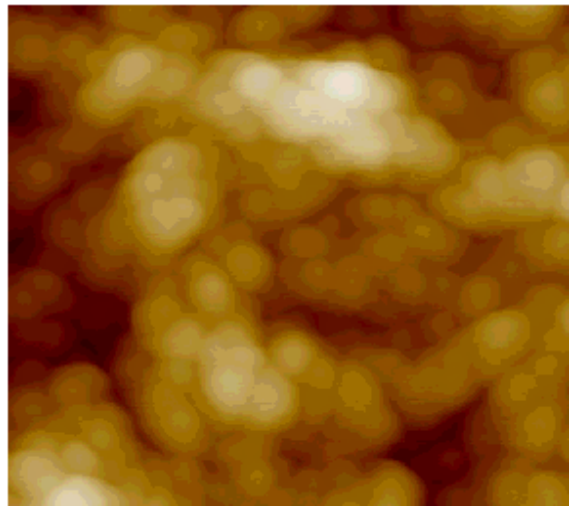

**He 250 nm**     **Ar 250 nm**

FIG. 9.  J.R. Rabeau

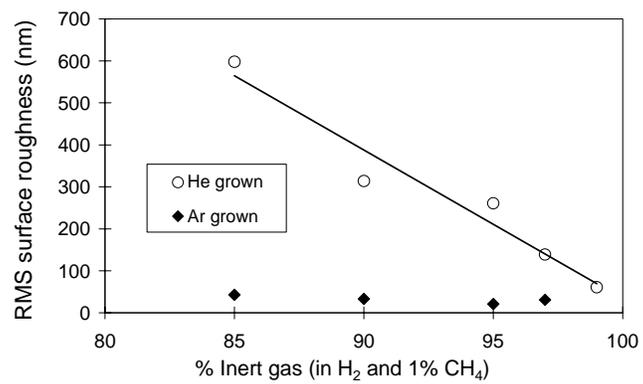

FIG. 10. J.R. Rabeau

(a)

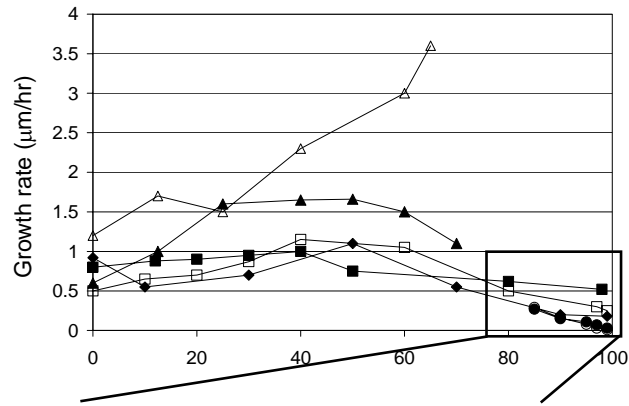

(b)

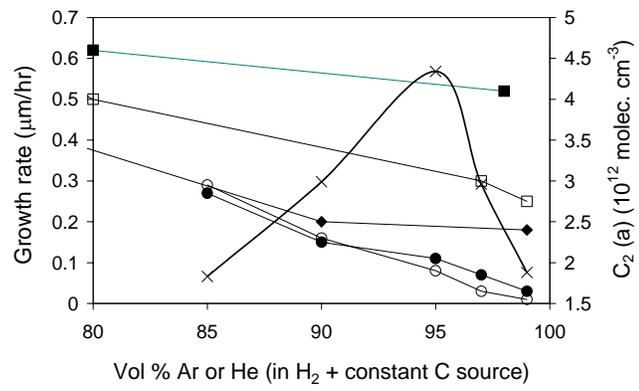

FIG. 11. J.R. Rabeau

**List of Figures**

FIG. 1. Coaxial blade reactor (showing only 2 of 6 diagnostic ports).

TABLE I. Growth conditions used in the CBR for growth of films with Ar and He.

FIG. 2. Typical OES spectra from (a) Ar/$H_2$/$CH_4$ and (b) He/ $H_2$/$CH_4$ plasmas in the coaxial blade reactor.

FIG. 3. $C_2$ (a) number density as a function of Ar content (1% $CH_4$ and a balance of $H_2$, 400 sccm total flow) in the CBR reactor at 2.1 kW and 90 Torr. The number density was measured using different spectral lines indicated by different symbols. The upper limit of $C_2$ (a) in the He containing plasmas is shown along the baseline of the plot.

FIG. 4. Raman spectra (244 nm) of samples grown with an increasing amount of Ar or He diluted $H_2$/$CH_4$ process gas compositions.

FIG. 5. Four Lorentzian curves (-) were fitted to the 97% Ar experimental Raman spectrum (•). The fit gave excellent agreement with the experimental data.

FIG. 6. Plot of the ratio of the diamond peak area to the 'graphite' peak area versus inert gas content for samples grown with an increasing amount of Ar or He diluted $H_2/CH_4$ process gas compositions.

FIG. 7. SEM micrographs of films grown with 97% Ar or He diluted $H_2/CH_4$ process gas compositions for 7.5 hours.

FIG. 8. The average growth rate of diamond films grown with an increasing amount of Ar or He diluted $H_2/CH_4$ process gas compositions. The error was estimated to be approximately 10%.

FIG. 9. AFM images of diamond films grown with 97% Ar or He diluted $H_2/CH_4$ process gas compositions for 7.5 hours.

FIG. 10. RMS roughness for films grown with an increasing amount of Ar or He diluted $H_2/CH_4$ process gas compositions.

FIG. 11. (a) Observed growth rates as a function of inert gas content (balance $H_2$ and constant $CH_4$ or $C_2H_5OH$ content) taken from the literature. ◆Yang et al.[36] Ar (900ºC), □Zhou et al.[5] Ar (800ºC), ΔBaranauskas et al.[38] Ar (855ºC), ▲Baranauskas et al.[37] He

(855°C), ■Han *et al.*[39] Ar (800 – 900°C), ○This work Ar (400 – 500°C), ●This work He (< 500°C). (b) shows a magnified portion of the plot and the $C_2$ number density measured in our experiments during Ar addition is superimposed for comparison.